\documentclass[twocolumn,showpacs,amsmath,amssymb]{revtex4}
\usepackage{graphicx}
\usepackage{dcolumn}
\usepackage{bm}
\usepackage{amsmath}
\usepackage{amsfonts}
\begin{document}

\title{Excluding Light Asymmetric Bosonic Dark Matter}
\author{Chris {\sc Kouvaris}}\email{kouvaris@cp3.sdu.dk}
\affiliation{$\text{CP}^3$-Origins, University of Southern Denmark, Campusvej 55, Odense 5230, Denmark}
\author{Peter {\sc Tinyakov}}\email{Petr.Tiniakov@ulb.ac.be}
 \affiliation{Service de Physique Th\'eorique,  Universit\'e Libre de Bruxelles, 1050 Brussels, Belgium}
 
\begin{abstract}
We argue that current neutron star observations exclude asymmetric
bosonic non-interacting dark matter in the range from 2~keV to 16~GeV,
including the 5-15~GeV range favored by DAMA and CoGeNT. If bosonic
WIMPs are composite of fermions, the same limits apply provided the
compositeness scale is higher than $\sim 10^{12}$ GeV (for WIMP mass
$\sim 1$~GeV). In case of repulsive self-interactions, we exclude large
range of WIMP masses and interaction cross sections which complements
the constraints imposed by observations of the Bullet Cluster.
\end{abstract}

\pacs{95.35.+d 95.30.Cq}

\maketitle 

{\it 1. Introduction.}  An appealing solution to the dark matter (DM)
problem is offered by Weakly Interacting Massive Particles (WIMPs)
emerging in many theories beyond the Standard Model (SM).  However,
WIMPs are very difficult to detect, and therefore little is known
about their properties. Experimentally, the situation is rather
unclear (see e.g. limits from CDMS~\cite{Ahmed:2009zw}), with
DAMA~\cite{Bernabei:2010mq} and CoGeNT~\cite{Aalseth:2010vx}
suggesting the existence of a light WIMP with a mass around $\sim
10$~GeV.

Apart from direct searches, constraints on WIMPs can be set by
observations of compact objects such as white dwarfs and neutron stars
\cite{Goldman:1989nd,Kouvaris:2007ay,Bertone:2007ae,Sandin:2008db,McCullough:2010ai,Kouvaris:2010vv,deLavallaz:2010wp,Kouvaris:2010jy,Ciarcelluti:2010ji}.
These constraints can be grouped in two types. The first type targets
WIMPs that can annihilate inside the star producing heat that can
change the thermal evolution of the star~\cite{Kouvaris:2007ay}.
WIMPs of this type can arise in supersymmetric extensions of the SM
(see \cite{Jungman:1995df} and references therein), or in Technicolor
models~\cite{Technicol}. Constraints of the second type target
asymmetric DM models. In these models the annihilation of DM in the
present-day Universe is impossible because only particles, and no
anti-particles (hence the term ``asymmetric'')
remain~\cite{Nussinov:1985xr,Hooper:2004dc,ldm1,Kaplan:2009ag,ldm2}.
%remain~\cite{Nussinov:1985xr,Hooper:2004dc,Gudnason:2006yj,Gudnason:2006ug,Khlopov:2007ic,Kouvaris:2008hc,Foadi:2008qv,Khlopov:2008ty,Sannino:2008wy,Kaplan:2009ag,Frandsen:2009mi,Hall:2010jx,Belyaev:2010kp,Taoso:2010zz,Buckley:2010ui,Dutta:2010va,Cohen:2010kn,Falkowski:2011xh,Frandsen:2011kt}.
An additional bonus in these models is that the asymmetry of WIMPs
might be linked through sphalerons with the baryon asymmetry
\cite{Hooper:2004dc,Kaplan:2009ag}, which can explain the today's
ratio $\Omega_{DM}/\Omega_{B}\sim 5$ provided the WIMP has a mass
around 5~GeV. Note that this value is not far from the one suggested
by DAMA and CoGeNT. In view of this coincidence, the models with WIMP
masses in the GeV range have become quite popular.

Since in the asymmetric DM models WIMPs cannot annihilate, if a large
number of them is accreted during the lifetime of a neutron star, they
may collapse forming a small black hole inside the star that
eventually destroy the latter. Therefore the existence of old neutron
stars can impose constraints on the properties of asymmetric WIMPs. In
fact, in the case of fermionic asymmetric WIMPs with a spin-dependent
cross section, these constraints are competitive to direct DM search
experiments~\cite{Kouvaris:2010jy}.

In this letter we focus on asymmetric bosonic dark matter and derive
constraints on the DM parameters from the formation of black holes
inside neutron stars. We show that for fundamental asymmetric
non-interacting bosonic WIMPs, current observational data exclude all
the masses from 2~keV to 16~GeV (including the range of masses
$5-15$~GeV favored by DAMA and CoGeNT). If WIMPs are composite
particles made of fundamental fermions, the above constraint does not
apply. However, if the compositeness scale is above $\sim 10^{12}$~GeV
(like in Grand Unified Theories (GUT)), candidates like that are again
excluded in the same mass range. In addition, we constrain the case of
fundamental self-interacting bosonic WIMPs. We show that if the
interaction is repulsive, a vast area of self-interaction cross
sections complementary to the one excluded by observations of the
Bullet Cluster~\cite{Randall:2007ph} is excluded.

{\it 2. Bosonic Dark Matter.} Gravitational collapse of a
self-gravitating lump of particles happens differently in case of
bosons and fermions. In the case of fermions of mass $m$, a large
number of particles $N \simeq (M_{\rm Pl}/m)^3$ is required to
overcome the Fermi pressure, where $M_{\rm {Pl}}$ is the Planck
mass. In the case of non-interacting bosons this number is
parametrically smaller, $N \simeq (2/\pi) (M_{\rm Pl}/m)^2$, since
only the uncertainty principle opposes the collapse. A repulsive 
interaction of bosons would provide an extra pressure, so the
required number of particles is larger in this case. Taking a
$\lambda \phi^4$ model as a generic example, the minimum mass of a
self-gravitating lump which can form a black hole
is~\cite{Mielke:2000mh}
\begin{equation}	 
M_{\rm crit}= \frac{2M_{\rm Pl}^2}{\pi m} \sqrt{1+
  \frac{ M_{\rm Pl}^2}{4\sqrt{\pi}  m} \sigma^{1/2}}
\label{m_crit} 
\end{equation} 
where we have expressed the result in terms of the self-interaction
cross section $\sigma = \lambda^2/(64\pi m^2)$. Here and below we use
the natural units $\hbar = c = k_B =1$.

It is easy to see from eq.~(\ref{m_crit}) that for cross sections
$\sigma \gg M_{\rm Pl}^4/m^2\sim 10^{-104}\,{\rm cm}^2 (m/{\rm
  GeV})^{-2}$, the second term dominates, and the minimum required
mass scales (at constant $\lambda$) in the same way as for the
fermionic particles with a different (and potentially much smaller)
coefficient. The best experimental constraints on the self-interaction
cross section come from the Bullet Cluster, $\sigma/m<2 \times
10^{-24}\text{cm}^2/\text{GeV}$~\cite{Randall:2007ph}.

Several conditions have to be satisfied for a gravitational collapse
of WIMPs inside a neutron star to occur. Firstly, a sufficient number
of DM particles must be accumulated during the lifetime of the neutron
star.  The accretion of WIMPs onto a typical 1.4$M_{\odot}$ neutron
star in a globular cluster, taking into account relativistic effects,
has been calculated in~\cite{Kouvaris:2010vv}. The total mass of
accreted WIMPs is
\begin{equation}	
M_{acc}=4.3 \times 10^{46} 
\left (\frac{\rho_{\text{dm}}}{10^3 \text{GeV}/\text{cm}^3} \right )  
\left (\frac{t}{\text{Gyr}} \right )f~\text{GeV}, 
\label{rate}  
\end{equation} 
where the ``efficiency'' factor $f=1$ if the WIMP-nucleon cross
section satisfies $\sigma_n > 10^{-45}\text{cm}^2$, and
$f=\sigma_n/(10^{-45}\text{cm}^2)$ if $\sigma_n <
10^{-45}\text{cm}^2$.  
The
condition
\begin{equation}
M_{acc}>M_{crit}
\label{eq:cond1}
\end{equation}
guarantees that the accumulated DM mass is above the critical
value (\ref{m_crit}). 

Secondly, the newly-formed black hole must accrete matter faster than
it evaporates due to Hawking radiation. In the Bondi regime of
accretion, the change of the black hole mass $M$ with time is given by
the equation
\begin{equation}
\frac{dM}{dt}=\frac{4\pi\rho_c G^2 M^2}{c_s^3}-\frac{1}{15360
\pi G^2 M^2}, \label{eq:dMdt}
\end{equation} 
where $c_s$ and $\rho_c$ are the speed of sound and the mass density
of the neutron star core, respectively. The first term corresponds to
the Bondi accretion while the second represents the energy loss due to
Hawking radiation. Since the accretion increases while the Hawking
radiation decreases as a function of $M$, it is the initial mass of
the black hole that determines its fate. Requiring that the first term
dominates when the black hole is formed gives the condition
\begin{equation}
M > 5.7 \times 10^{36}~\text{GeV}.
\label{eq:hawking}
\end{equation}
Here we have used $\rho_c = 5\times 10^{38}~\text{GeV}/\text{cm}^3$
and $c_s=0.17$. Any black hole with the initial mass satisfying
eq.~(\ref{eq:hawking}) will eventually destroy the whole star, while
the smaller black holes will evaporate with no detectable effect.

The third condition necessary for the WIMP collapse into a black hole
is the onset of the WIMP self-gravitation. WIMPs captured by the
neutron star thermalize within a time $t_{\rm th} = 2\times
10^{-5}{\rm yr}\,(m/{\rm
  GeV})^2$~\cite{Goldman:1989nd,Bertone:2007ae,Kouvaris:2010vv} and
concentrate in the center within the radius
\begin{equation}
r_{\rm th} \simeq 2 ~\text{m} 
\left( {T_c\over 10^5 {\rm K}} \right)^{1/2}
\left( {m\over {\rm GeV}} \right)^{-1/2},
\label{eq:rth}
\end{equation} 
where $T_c$ is the temperature of the star core.  When their total
mass $M$ increases beyond the mass of the ordinary matter within the
same radius,
\begin{equation}
M > \frac{4}{3}\pi \rho_c r_{\rm th}^3 = 2.2 \times
10^{46}~\text{GeV}
\left( {m\over {\rm GeV}} \right)^{-3/2},
\label{eq:selfgrav}
\end{equation}
their own gravitational field starts to dominate over that of the star
and the self-gravitation regime sets in, leading to the gravitational
collapse provided the condition (\ref{eq:cond1}) is satisfied. It can
be seen from eq.~(\ref{rate}) that (\ref{eq:selfgrav}) is satisfied if
the WIMP mass is larger than $\sim$1 GeV ($\sim$143 GeV) for
$\rho_{\rm dm}=10^3 \text{GeV}/\text{cm}^3$ ($\rho_{\rm dm}=0.3
\text{GeV}/\text{cm}^3$), but not for lighter WIMPs.

However, if WIMPs are bosons they can form a Bose-Einstein condensate
(BEC). Since this state is more compact, the self-gravitation in this
case starts for a smaller number of particles, i.e., before the
condition (\ref{eq:selfgrav}) is satisfied.  The particle density
required to form BEC is
\[
n \simeq  4.7\times 10^{28} \text{cm}^{-3} 
\left( {m\over {\rm GeV}} \right)^{3/2} 
\left (\frac{T_c}{10^5 \text{K}} \right )^{3/2}.
\]
Assuming an old neutron star with a temperature $T_c=10^5$~K, the
number of WIMPs needed in order for BEC to form is $N_{\rm BEC} \simeq
2 \times 10^{36}$.  All the WIMPs accreted in excess of this value
will go into the condensed state. For most of the cases of our
interest, the number of accreted WIMPs will be larger than $N_{\rm
  BEC}$, so eq.~(\ref{eq:selfgrav}) has to be reconsidered.

The size of the condensed state is determined by the radius of the
wave function of the WIMP ground state in the gravitational potential of
the star,
\begin{equation}
r_c = \left (
\frac{8\pi }{3} G \rho_c m^2\right )^{-1/4}\simeq 1.6 \times 10^{-4} \left
(\frac{\text{GeV}}{m} \right )^{1/2}
\text{cm}.
\label{condensed_ground}
\end{equation}
Substituting this size in place of $r_{\rm th}$ in
eq.~(\ref{eq:selfgrav}) we get 
\begin{equation}
M > 8\times 10^{27}~\text{GeV} 
\left( {m\over {\rm GeV}} \right)^{-3/2}.
\label{eq:}
\end{equation}
In view of eq.~(\ref{rate}), the amount of DM sufficient for WIMP
self-gravitation in the condensed state can always be accumulated
provided that the WIMP is heavier than $\sim 0.1$~eV, which covers all
cases of interest. Thus, due to the formation of BEC the requirement
of self-gravitation does not provide an extra condition.

Finally, the accumulation of WIMPs may become inefficient if they may
escape from the neutron star once captured, which is a danger at small
WIMP masses. It can be seen from eq.~(\ref{eq:rth}) that for WIMP
masses in the keV range the radius of the WIMP lump becomes comparable
to the size of the star, so that WIMPs in the tail of the velocity
distribution may escape.  The rate $F$ of WIMP evaporation can be
estimated as follows~\cite{Krauss:1985aaa},
\begin{equation}
F=n_s\left( {T\over 2\pi m} \right)^{1/2}
\left(1+{GMm\over RT} \right)\exp\left(-{GMm\over RT}\right), 
\end{equation}
where $T$, $M$ and $R$ are the temperature, mass and radius of the
star, respectively, and $n_s$ is the WIMP density at the
surface. Calculating $n_s$ from the Boltzmann distribution we found
that the evaporation can be safely ignored for masses larger than
$\sim 2$~keV.

\begin{figure}[!tbp]
\includegraphics[width=1.0\linewidth,height=0.65\linewidth]{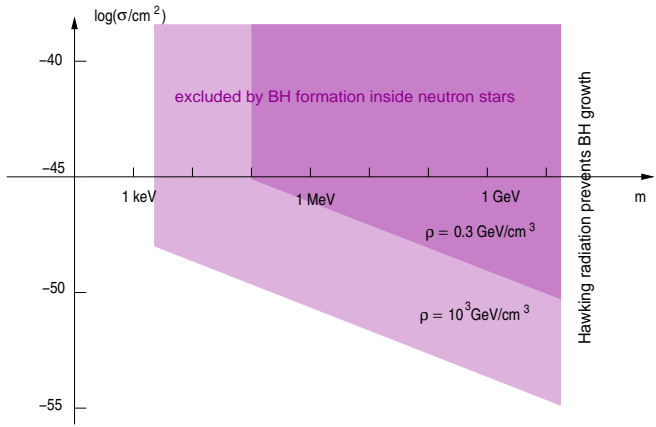}
\caption{The exclusion regions as a function of the DM mass and the
  WIMP-to-nucleon cross section for an isolated neutron star at local
  DM density $\rho = 0.3\,$GeV/cm$^3$ (such as J0437-4715 and
  J0108-1431) and for a neutron star in the core of a globular cluster
  at $\rho = 10^3\,$GeV/cm$^3$.
\label{fig:1}
}
\end{figure}

In summary, accumulation and subsequent gravitational collapse of
WIMPs captured inside a neutron star occur for WIMPs heavier than
$\sim 2$~keV provided the conditions (\ref{eq:cond1}) and
(\ref{eq:hawking}) are satisfied. In the case of no self-interactions
the collapse to a black hole inside the neutron star happens for WIMP
masses $2~\text{keV} \lesssim m \lesssim 16~\text{GeV}$. The upper
bound of this exclusion range is independent of the local DM density
$\rho$, while the lower bound raises slightly at small $\rho$. As a
function of the DM mass, the bound on the WIMP-to-nucleon cross
section is $\sigma_n > 8\times 10^{-50}\text{cm}^2({\rm GeV}/m)$ for nearby
isolated stars at local
  DM density $\rho = 0.3\,$GeV/cm$^3$ 
and $\sigma_n > 2\times 10^{-54}\text{cm}^2({\rm
  GeV}/m)$ for stars in globular clusters at local
  DM density $\rho = 10^3\,$GeV/cm$^3$. The resulting exclusion
regions are shown in Fig.~\ref{fig:1}. 

Several old neutron stars have been observed, both in the vicinity of
the Earth where the DM density is $\rho \sim 0.3\,$GeV/cm$^3$, and in
the cores of globular clusters where the DM density may be as high as
$10^3-10^4\,$GeV/cm$^3$ \cite{McCullough:2010ai}.  Examples of nearby
neutron stars are J0437-4715~\cite{Kargaltsev:2003eb} and
J0108-1431~\cite{Mignani:2008jr} (140~pc and 130~pc from the Earth,
respectively). The examples of neutron stars in globular clusters are
e.g. the pulsar B1620-26 located at the outskirts of the core of M4,
and X7 from 47 Tuc~\cite{Rybicki:2005id}.

\begin{figure}[!tbp]
\begin{center}
\includegraphics[width=1.0\linewidth,height=0.65\linewidth]{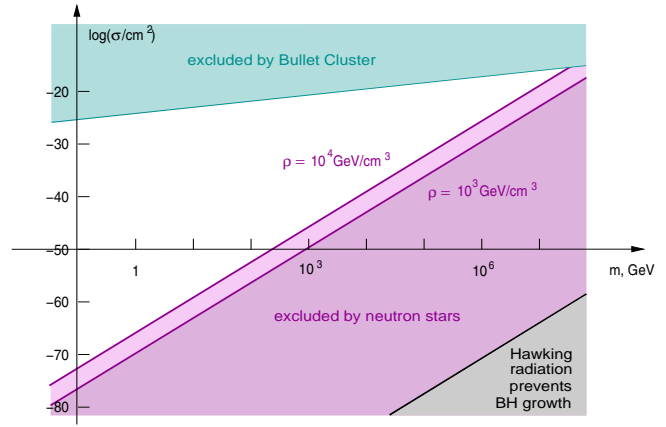}
\caption{Constraints on bosonic DM mass $m$ and self-interaction cross
  section $\sigma$ from observations of neutron stars in globular
  clusters. Excluded region (pink) is shown for two background DM
  densities as indicated on the plot. The cyan region shows
  constraints from the Bullet Cluster.
\label{fig:2}
  }
\end{center}
\end{figure}

In the case of a repulsive interaction, the exclusion region that
follows from eqs.~(\ref{eq:cond1}) and (\ref{eq:hawking}) is shown in
Fig.~\ref{fig:2}. Depending on the self-interaction cross section, the
constraints extend to much higher masses and are complementary to
those derived from the observation of the Bullet Cluster.

{\it 3. Composite Dark Matter.} The discussion above refers
specifically to fundamental bosonic DM. If instead the latter is
composite of fermions, the situation might change.  There are two
possibilities. Assume the total WIMP mass exceeds $M_{\rm crit}$,
eq.~(\ref{m_crit}). As the DM lump shrinks towards its Schwarzschild
radius, it might reach the density at which the mean distance between
WIMPs is comparable to the scale of compositeness.  At this point the
Fermi pressure comes into play and might stop further collapse unless
the DM lump has already reached its Schwarzschild radius.

To estimate the minimum compositeness scale $\Lambda_{\rm crit}$, we
express the mean distance $d$ between WIMPs in the nearly-collapsing
(i.e., having size comparable to its Schwarzschild radius) DM lump in
terms of its mass $M$. Ignoring the numerical coefficients, we have
$d=GM^{2/3}m^{1/3}$. Taking the mass to be equal to the critical one,
eq.~(\ref{m_crit}), we get
\begin{equation}
\Lambda_{crit}=m^{1/3}M_{\rm Pl}^{2/3} 
\left(1+ \frac{ \lambda m_{pl}^2}{32 \pi m^2} \right)^{-1/3}.
\end{equation} 
In the non-interacting case this gives $\Lambda_{crit}= 2 \times
10^{12}~\text{GeV} (m/\text{GeV})^{1/3}$, which is well below the GUT
mass scale of order $\sim 10^{16}$~GeV for all masses of interest
(cf. Fig.~\ref{fig:2}). Thus, our constraints are also valid for
composite WIMPs with the compositeness scale higher than $\sim
10^{12}$~GeV. 

{\it 4. Discussion and conclusions.}  Two remarks are in order. In the
above analysis we have assumed that the black hole that is formed
inside a neutron star and is not destroyed by the Hawking radiation
eventually consumes the whole star. However plausible, this assumption
requires some justification. In eq.~(\ref{eq:dMdt}) we have taken
Bondi accretion, which is very efficient and would indeed consume the
whole star much faster than in 1~Gyr (see Ref.~\cite{Kouvaris:2010jy}
for the calculation). However, the accretion may be slowed down by the
angular momentum of the star. The Bondi regime cannot be maintained if
the angular momentum of the falling matter exceeds the one it would
have on the innermost stable orbit. One can show that, in the absence
of momentum transfer, for a typical period of an old neutron star of
the order of a second, only a small inner part of the star can be
consumed in the Bondi regime. The momentum transfer has been studied
in Ref.~\cite{Markovic:1994bu} in the context of ordinary
stars. Rescaling the parameters to the case of a neutron star, we
found that if the momentum transfer due to the viscosity is taken into
account, the Bondi accretion is maintained until the black hole
reaches a mass of $\sim 10^{-7} M_{\odot}$. From this state, with the
Bondi rate the consumption of the whole star would take about 1~min.
Even if the actual rate is many orders of magnitude slower, the star
will definitely be destroyed within 1~Gyr.

The second remark concerns the validity of the Bondi regime at the
initial stages of the black hole growth. The lightest black hole
relevant for our analysis has size and Hawking temperature in the
GeV range. While the fluid approximation should still be
adequate at these scales because of the extremely high density of the
nuclear matter, the actual parameters may differ from those used in
eq.~(\ref{eq:dMdt}). This might change slightly the upper value of the
exclusion mass range. The precise calculation is difficult and goes
beyond the scope of this letter. 

To conclude, we have demonstrated that the existing observations of
old neutron stars, both in globular clusters and in the vicinity of
the Earth, exclude light non-interacting fundamental bosonic dark
matter candidates in the mass range from 2~keV to about 16~GeV with
WIMP-to-nucleon cross section $\sigma_n > 8\times
10^{-50}\text{cm}^2({\rm GeV}/m)$. The constraints equally apply to
composite bosonic dark matter if the compositeness scale is higher
than $\sim 10^{12}$~GeV. A wide range of masses and cross sections is
also excluded for very weakly self-interacting bosonic candidates.

The work of P.T. is supported by the IISN project No. 4.4509.10 and by
the ARC project ``Beyond Einstein: fundamental aspects of gravitational
interactions''.

{\bf Note added:} When this paper was being finalized
Ref.~\cite{McDermott:2011jp} appeared which addressed the same
question. However, in Ref.~\cite{McDermott:2011jp} the Hawking
evaporation of black holes was disregarded which invalidates the
constraints for WIMP masses $m>16$~GeV. At smaller masses our results
are somewhat different because the effect of the Pauli blocking was
overestimated in Ref.~\cite{McDermott:2011jp}.

\end{document}